\documentstyle[aps,prl,epsf]{revtex}

\begin{document}

\title{Recovery of the Linearized 4D AdS and dS Schwarzschild Metric in the
Karch-Randall Braneworld}

\author{Ioannis Giannakis${}^{a}$
\footnote{E-mail: giannak@summit.rockefeller.edu}}
\address{
${}^{(a)}$ Physics Department, The Rockefeller University\\
1230 York Avenue, New York, NY 10021-6399}

\maketitle

\begin{abstract}
We present a linearized treatment of the Karch-Randall braneworld
where an AdS$_4$ or dS$_4$ brane is embedded in AdS$_5$. We examine
the quasi-zero graviton mode in detail and reproduce the graviton
mass by elementary means for the AdS$_5$ case.
We also determine the axially symmetric,
static excitations of the vacuum and demonstrate that they
reproduce the 4D AdS$_4$ and dS$_4$ Schwarzschild metrics on
the brane.
\end{abstract}


\ifpreprintsty\else
\fi


In this paper we discuss black holes in the
braneworld \cite{randall}, \cite{karch} in connection
with the series of our
earlier publications \cite{liu}, \cite{ren},
\cite{ioannis}.
The physical brane in the Karch-Randall braneworld is a
4D hyperface embedded in a 5D manifold that is asymptotically AdS.
The Gauss-normal form of the
metric in this space is
\begin{equation}
ds^2=e^{2A(y)}{\bar g}_{\mu\nu}(x,y)dx^\mu dx^\nu+dy^2,
\label{eqena}
\end{equation}
which is the most general metric with four-dimensional covariance.
Here and throughout the
paper, we adopt the convention that the Greek indices take values
0--3 and the Latin indices 0--4.  Thus $g_{mn}$ is the full
five-dimensional metric, while $\bar g_{\mu\nu}$ is the four-dimensional
metric ``on the brane''.

A simple solution for the warp factor $A(y)$ was proposed for the case
of a brane with negative or positive cosmological
constant embedded in an AdS$_5$ bulk \cite{nihei}, \cite{kaloper},
\cite{kim}, \cite{dewolfe}, \cite{karch}
\begin{equation}
e^{2A} = \left({k\over \kappa}\right)^2{\cosh^2{\kappa(|y|-y_0)}},
\qquad
e^{2A} = \left({k\over \kappa}\right)^2{\sinh^2{\kappa(|y|-y_0)}},
\label{eqdisitter}
\end{equation}
where $y_0$ is a constant related to the value of the warp factor on the
brane (located at $y=0$),
$\Lambda_5=-4\kappa^2$ and $\Lambda_4=\pm3k^2$.
Note that we have inserted an
absolute value in the $y$ coordinate, which enforces the location of the
brane.  This discontinuity in the slope of the warp factor can be fixed
by the Israel matching conditions that account for the brane tension.
In this paper we concentrate on the
AdS$_4$ brane and comment
on the dS$_4$ brane only in case of qualitative differences
with the treatment of the AdS$_4$.

We treat the above solutions as the braneworld vacua, and consider
the effect of an axially symmetric perturbation about these
backgrounds.  By noting that the AdS$_4$ metric,
$\bar g_{\mu\nu}$,
may be written in global coordinates as
\begin{equation}
\bar d\bar s^2=-(1+k^2r^2)dt^2
+\frac{dr^2}{1+k^2r^2}+r^2d\Omega^2.
\label{eqglobads}
\end{equation}
we take the most general axially
symmetric and static metric in $D=4+1$ with the following form
\begin{equation}
ds^2=e^{2A}[-e^{a+{\bar a}}dt^2
+e^{b-{\bar a}}dr^2+e^{c}r^2d\Omega_2^2]+dy^2,
\label{eqpente}
\end{equation}
where $e^{2A}$, given by Eq.~(\ref{eqdisitter}), and
${\bar a}={\ln({1+k^2r^2})}$ are fixed background quantities.
This general ansatz depends on three
functions $a$, $b$ and $c$, all functions of $r$ and $y$.
Substituting the metric Eq.~(\ref{eqpente}) into the Einstein equations
and keeping terms linear in $a$, $b$ and $c$, we obtain
the following equations
\begin{eqnarray}
& e^{\bar a} & \Big[a^{\prime\prime}
+\frac{2}{r}a^\prime +\frac{1}{2}{\bar a}^\prime(3a'-b'+2c')\Big]
+e^{2A}\Big[\ddot a+ \dot A(5\dot a+\dot b+2\dot c)\Big]
=6k^2b \nonumber\\
& e^{\bar a} & \Bigl[a^{\prime\prime}+2c^{\prime\prime}-\frac{2}{r}(b'-2c')
+\frac{1}{2}{\bar a}^\prime(3a^\prime-b^\prime+2c^\prime)\Bigr]
+e^{2A}\Bigl[\ddot b+\dot A(\dot a+5\dot b+2\dot c)\Bigr]
=6k^2 b \nonumber\\
& e^{\bar a} & \Bigl[c^{\prime\prime}+\frac{1}{r}(a'-b'+4c')
+{\bar a}^{\prime}c^\prime \Bigr]+
e^{2A}\Bigl[\ddot c+\dot A(\dot a+\dot b+6\dot c) \Bigr]
-\frac{2}{r^2}(b-c)
=6k^2b \nonumber\\
& \ddot a & +\ddot b+2\ddot c+2\dot A(\dot a
+\dot b+2\dot c)=0, \nonumber\\
& \dot a^\prime & +2\dot c^\prime-\frac{2}{r}(\dot b-\dot c)
+\frac{1}{2}{\bar a}^\prime(\dot a-\dot b)=0,
\label{eqenea}
\end{eqnarray}
where prime and dot denote differentiation with respect
to $r$ and $y$.  Note that we have omitted a brane
stress energy tensor, $T^\mu_\nu\delta(y)$, as a source. 
The interested reader should consult reference \cite{liu}
for a detailed analysis in the presence of matter on the brane.
The brane tension is accounted
for by the kink in the warp factor, Eq.~(\ref{eqdisitter}).
By integrating
Eq.~(\ref{eqenea}) across the brane, one obtains the Israel
matching conditions
on the metric functions $a$, $b$ and $c$.  In the absense of sources
on the brane the Israel matching conditions become the
Neumann boundary conditions.
These equations constitute the starting point of our
investigation.

The gauge symmetries of these equations,
the coordinate transformations that respect the
axially symmetric, static form of the metric, Eq.~(\ref{eqpente}),
are generated by the functions
$v$ and $u$ that obey the following relations to linear order
\begin{equation}
e^{2A-{\bar a}}{\dot v}+u^\prime=0, \qquad
{\dot u}=0.
\label{eqreann}
\end{equation}
The constraints, (\ref{eqreann}), may be
integrated to provide the following
form for the parameters of transformation
\begin{equation}
u=\chi(r), \qquad v(r, y)=-e^{\bar a}{\chi}^{\prime}{\int^y}d{x}\,
e^{-2A(x)}+{\phi}(r).
\label{eqriot}
\end{equation}
Consequently, the axially symmetric gauge transformations
may be fully parameterized
by two functions of $r$, namely $\chi(r)$ and $\phi(r)$.

The solution to the linearized equations of motion should satisfy the
Neumann boundary condition $\dot a(r, 0)=\dot b(r, 0)=\dot c(r, 0)$
on the physical brane and the asymptotic conditions
\begin{equation}
\lim_{|y|, r\to\infty}a=\lim_{|y|, r\to\infty}b
=\lim_{|y|, r\to\infty}c=0.
\label{eqexy}
\end{equation}
The penultimate
equation of (\ref{eqenea}),
corresponding to $R_{yy}$, may be integrated to give
\begin{equation}
a+b+2c=0.
\label{eqvirile}
\end{equation}
By substituting
$c=-{1\over 2}(a+b)$ into the final equation of (\ref{eqenea}) and
integrating, we find
\begin{equation}
b^\prime+\frac{1}{r}(a+3b)-\frac{1}{2}{\bar a}^\prime(a-b)=0.
\label{eqrichos}
\end{equation}
Furthermore, by taking into account that ${\bar a}=
{\ln(1+k^2r^2)}$, we eliminate $a$ in terms of $b$
\begin{equation}
a=b-(1+k^2r^2)(rb^\prime+4b).
\label{eqriotx}
\end{equation}
This allows us to rewrite the equations of motion in terms of the single
function $b(r,y)$.  We now separate variables for the
transverse and traceless modes by decomposing
$b(r, y)={\sum_\mu}B(\mu)b(r|\mu){\psi}(y|\mu)$.
By separating variables on the second equation
of (\ref{eqenea}) we derive
a second order differential equation for the modes $b(r|\mu)$
\begin{equation}
b^{\prime\prime}+\frac{4}{r}\frac{1+2k^2r^2}{1+k^2r^2}
b^\prime
+\frac{10k^2-{\mu}^2}{1+k^2r^2}b=0,
\label{eqvios}
\end{equation}
as well as an eigenvalue equation for the modes ${\psi}(y|\mu)$
\begin{equation}
e^{2A(y)}({\ddot{\psi}}+4{\dot A}{\dot{\psi}})=
-{\mu}^2{\psi}.
\label{eqwilde}
\end{equation}

Eq.~(\ref{eqwilde}) defines a Sturm-Liouville problem with $e^{2A}$ 
as the measure of the normalization integral. Furthermore,
by imposing the boundary 
conditions $e^{2A}\psi\to 0$ as $y\to\infty$ and $\dot\psi=0$ at 
$y=0$, we have a normalization integral
\begin{equation}
\int_0^\infty dy\,e^{2A(y)}\psi^2(y|\mu)=\psi(0|\mu)\frac{\partial\dot
\psi(0|\mu)}{\partial\mu^2}.
\label{eqconst}
\end{equation}

It was demonstrated in \cite{karch} that the graviton spectrum in the
AdS (dS) braneworld is composed of a quasi-zero (zero)
mode graviton trapped on the
brane as well as a discrete tower of Kaluza-Klein modes.  This spectrum
may be determined by examining the
eigenvalue equation (\ref{eqwilde})
for the graviton wavefunction.  In this paper, we examine the quasi-zero
mode and reproduce the expression for the
graviton mass
by elementary means.  The method that we employ does not depend on the
details of the warp factor $A(y)$. Furthermore we
calculate the contribution of the quasi-zero mode ( zero mode in
the case of dS$_4$ brane ) to the axially symmetric, static metric
which when confined on the brane reproduces the AdS$_4$
(dS$_4$) Schwarzschild metric. The contributions due to
the radion mode and
the Kaluza-Klein tower have appeared in \cite{liu}.
The unexpected appearance
of the radion in the spectrum of the AdS braneworld
agrees with the result \cite{chatko}, which showed that
the radion in a two AdS brane model survives in the limit
in which the second brane is removed.

We begin by assuming that the mode
equation, (\ref{eqwilde}), admits a normalized
solution with eigenvalue $\mu^2 \ll k^2$, where $k^2$ is a scale introduced
by the curvature of the brane.
In this case
the right hand side of the equation can be treated perturbatively.
To zeroth order, the right hand side of (\ref{eqwilde}) is absent, and
the equation possesses two solutions
\begin{equation}
u_1=1, \qquad u_2={\tanh{\eta}}\,(2+{\rm sech}^2{\eta}),
\label{eqhakan}
\end{equation}
where $\eta\equiv \kappa(y-y_0)$. As a result, the normalizable zeroth
order solution is given by
\begin{equation}
\psi^{(0)}=1-\frac{1}{2}{\tanh{\eta}}\,(2+{\rm sech}^2{\eta}).
\label{eqzeroth}
\end{equation}
This solution fails to satisfy the Israel matching condition.
The first order correction
to the solution, $\psi^{(1)}(y)$, is determined by the inhomogeneous equation
\begin{equation}
{\ddot\psi^{(1)}}+4{\tanh{\eta}}{\dot\psi^{(1)}}
=-{\epsilon}{\rm sech}^2{\eta}\psi^{(0)}.
\label{eqboulez}
\end{equation}
where $\epsilon=\frac{\mu^2}{k^2}$.
This equation may be solved by the method of variation of coefficients;
the solution is given by
$\psi^{(1)}=\mu^2(c_1u_1+c_2u_2)$, where $c_1$ and $c_2$ are both
functions of $\eta$ and are given by the expressions
\begin{equation}
c_1=-\frac{1}{3k^2}{\int_{\eta}^\infty }d\zeta\,
{\psi^{(0)}}(\zeta)(2{\cosh^{2}{\zeta}}+1){\tanh{\zeta}},
\qquad c_2=\frac{1}{3k^2}\int_{\eta}^\infty d{\zeta}\,
{\psi}^{(0)}(\zeta){\cosh^{2}{\zeta}}.
\label{eqrioz}
\end{equation}

Having carried out the integrations, we found
that the quasi-zero mode wavefunction
is given to first order in $\frac{\mu^2}{\kappa^2}$ as
\begin{eqnarray}
\psi={\psi^{(0)}}+\mu^2(c_1u_1+c_2u_2)
=1-\frac{1}{2}{\tanh{\eta}}\,(2+{\rm sech^2}{\eta})
+\frac{\mu^2}{3k^2}
\Bigl[\frac{1}{2}{\rm sech}^2\eta-2(1-\tanh^2\eta) \nonumber\\
+\frac{1}{6}\left(1-\frac{1}{2}\tanh\eta\,
(2+{\rm sech}^2\eta)\right)
+\ln(1+e^{-2\eta})\left(1
+\frac{1}{2}\tanh\eta(2+{\rm sech}^2\eta)\right)
\Bigr].
\label{eqroustou}
\end{eqnarray}
The four-dimensional mass parameter $\mu$ is calculated
by demanding that the quasi-zero mode wavefunction satisfies
the Israel matching condition:
\begin{equation}
\mu^2 \approx 6k^2e^{-2{\kappa}{y_0}} = \frac{3}{2}
\frac{k^4}{\kappa^2},
\label{eqvint}
\end{equation}
provided $\mu^2/k^2\ll1$.

Having identified the quasi-zero mode
wavefunction, (\ref{eqroustou}), and
the graviton mass, (\ref{eqvint}), we now proceed to determine the profile of
$b(r)$ on the brane.  This is accomplished by finding solutions of the
AdS$_4$ wave equation, (\ref{eqvios}), consistent with boundary conditions.
By introducing a new variable $\xi=-k^2r^2$, and defining
$b=f/(1-\xi)$, Eq.~(\ref{eqvios}) may be transformed to the
form
\begin{equation}
\frac{{\partial^2}}{{\partial{\xi^2}}}f+\frac{5}{2\xi}
\frac{{\partial}}{{\partial\xi}}f=-\frac{{\overline\epsilon}}
{{\xi}(1-\xi)}f,
\label{eqpertu}
\end{equation}
where $\overline\epsilon=\mu^2/4k^2$.

To zeroth order, there are simply two linearly independent solutions
\begin{equation}
u_1=1, \qquad u_2=(-\xi)^{-\frac{3}{2}}.
\label{eqcrios}
\end{equation}
Imposing boundary conditions at spatial infinity, we demand for the
zeroth order solution that $b^{(0)}\to0$ as $\xi\to-\infty$.  This
ensures that the spacetime remains asymptotically AdS$_4$ on the brane.
This requirement selects the second solution, so that we take
\begin{equation}
f^{(0)}=C(-\xi)^{-\frac{3}{2}}=\frac{C}{k^3r^3},
\label{eqri}
\end{equation}
where $C$ is a constant.
The next order solution can be written as
\begin{equation}
f^{(1)}=-\frac{4\overline\epsilon C}{3}
\left[(-\xi)^{-1/2}-\tan^{-1}(-\xi)^{-1/2}
+\frac{1}{2}(-\xi)^{-3/2}\ln(1-\xi)\right].
\label{eqvo}
\end{equation}
As a result, $b(r)$ is given up to first order in
$\overline\epsilon\approx3k^2/8\kappa^2$ by
\begin{equation}
b(r)=\frac{C}{{1+k^2r^2}}\left[\frac{1}
{{k^3r^3}}\left(1-\frac{k^2}{4\kappa^2}\ln(1+k^2r^2)\right)
-\frac{k^2}{2\kappa^2}\left(\frac{1}{kr}-\tan^{-1}\frac{1}{kr}\right)
\right].
\label{eqpoil}
\end{equation}
The remaining functions, $a(r)$ and $c(r)$ may now be determined through
Eq.~(\ref{eqriotx}) and the relation $c=-\frac{1}{2}(a+b)$.  The result is
\begin{eqnarray}
a(r) & = & \frac{C}{{1+k^2r^2}}\left[\frac{1}{kr}\left(1-\frac{k^2}
{4{\kappa}^2}
\ln(1+k^2r^2)\right)+\frac{k^2}{2\kappa^2}
(3+2k^2r^2)\left(\frac{1}{kr}-\tan^{-1}
\frac{1}{kr}\right)\right], \nonumber\\
c(r)& = & -\frac{C}{2}\left[\frac{1}{{k^3r^3}}\left(1-
\frac{k^2}{4\kappa^2}
\ln(1+k^2r^2)\right)+\frac{k^2}{\kappa^2}\left(
\frac{1}{kr}-{\tan^{-1}\frac{1}{kr}}\right)\right].
\label{eqcuio}
\end{eqnarray}

Given the quasi-zero mode graviton, this above solution ought to
reproduce the Schwarzschild-AdS black hole at linearized
order.  However, in order to compare this metric, computed in the
transverse-traceless gauge, with that in the standard Schwarzschild-AdS
form, we must transform the metric component $c$ to zero.  This may be
accomplished by making use of the residual coordinate
transformations,
(\ref{eqriot}), with parameter $\phi(r)$
\begin{equation}
a \to a+\frac{2k^2r}{1+k^2r^2}\phi(r), \qquad
b \to b-\frac{2k^2r}{1+k^2r^2}\phi(r)+2{\phi}^{\prime}(r),
\qquad c \to c+\frac{2}{r}{\phi}(r).
\label{eqdiop}
\end{equation}
We see that $c$ may be eliminated by choosing $\phi(r)
=-\frac{rc(r)}{2}$,
whereupon the transformed metric components take the form
\begin{eqnarray}
a(r) & = & -\frac{2G_4M}{r{(1+k^2r^2)}}\left[1
-\frac{k^2}{4\kappa^2}\ln(1+k^2r^2)+\frac{k^3r}{\kappa^2}
(1+k^2r^2)\left(\frac{1}{kr}
-\tan^{-1}\frac{1}{kr}\right)\right], \nonumber\\
b(r) & = & \frac{2G_4M}{r{(1+k^2r^2)}}\left[1+\frac{k^2}{2\kappa^2}
-\frac{k^2}{4\kappa^2}\ln(1+k^2r^2)\right].
\label{eqpoilas}
\end{eqnarray}
Here we have identified the dimensionless constant $C$ with the mass
according to $C=-4G_4Mk/3$.  This ensures that, up to terms of
${\cal O}(k^2/\kappa^2)$, this solution reproduces the linearized portion
of Schwarzschild-AdS
\begin{equation}
ds^2=-\left(1-\frac{2G_4M}{r}+k^2r^2\right)dt^2
+\left(1-\frac{2G_4M}{r}+k^2r^2\right)^{-1}dr^2+r^2d\Omega^2.
\label{eqwest}
\end{equation}
The normalization of $C$ was obtained in \cite{liu}, by considering
an explicit source on the brane.

This demonstrates that the quasi-zero mode graviton is responsible for
the long-range gravitational interaction on the brane, even though it
has mass $\mu^2\sim k^4/\kappa^2$.  This is a feature of the AdS$_4$
geometry in that, for the limit we are considering, the mass is
infinitesimal compared to the natural scale of the AdS curvature
({\it i.e.}~$\overline\epsilon\ll1$ for the dimensionless mass
$\overline\epsilon$).
Because of this, the background curvature provides a cutoff to the space
before any effects of the graviton mass may be discerned.  Furthermore,
we observe that no van Dam-Veltman-Zakharov
discontinuity arises \cite{dam}, \cite{zachkarov},
which is consistent with the results
\cite{kogan}, \cite{porrati}.

Let's now briefly discuss the dS$_4$ brane embedded in
the AdS$_5$ bulk. Corrections to the Newton's law in this
framework were discussed in \cite{kehagias}, \cite{ito}, \cite{nojiri},
\cite{kazuo}.
The equations of motion for the
modes $b(r|\mu)$ and $\psi(y|\mu)$ result from Eq.~(\ref{eqvios})
and Eq.~(\ref{eqwilde}) with the substitution
$k \mapsto ik$. To zeroth order Eq.~(\ref{eqwilde}) possesses
two linearly independent solutions
\begin{equation}
u_1=1, \qquad u_2={\coth{\eta}}\,(2-{\rm csch}^2{\eta}).
\label{eqzero}
\end{equation}
The constant solution is normalizable and
satisfies the Israel matching condition.
Furthermore we can also determine the profile of
$a(r)$, $b(r)$ and $c(r)$ on the brane. We find that
\begin{equation}
a(r)=\frac{C}{kr(1-k^2r^2)}, \qquad
b(r)=-\frac{C}{k^3r^3(1-k^2r^2)}, \qquad c(r)=\frac{C}{2k^3r^3}. 
\label{eqre}
\end{equation}
By performing a residual coordinate transformation
generated by $\phi(r)=-\frac{rc(r)}{2}$ we recover
the metric on the brane
\begin{equation}
a(r)=\frac{3C}{2kr(1-k^2r^2)}, \qquad
b(r)=-\frac{3C}{kr(1-k^2r^2)},\qquad c(r)=0. 
\label{eqren}
\end{equation}
which can be put into the standard Schwarzschild dS form
by identifying $C=-\frac{4G_4Mk}{3}$.
Eq.~(\ref{eqren}) coincides with the linearized approximation
of the Schwarzschild-dS metric
\begin{equation}
ds^2=-\left(1-\frac{2G_4M}{r}-k^2r^2\right)dt^2
+\left(1-\frac{2G_4M}{r}-k^2r^2\right)^{-1}dr^2+r^2d\Omega^2.
\label{eqeast}
\end{equation}
Let's recapitulate what we have done in this paper. We presented
a linearized analysis of the Karch-Randall model. In the
case of the AdS$_4$ (dS$_4$) brane there is a quasi-zero (zero)
mode in the spectrum. We calculated the mass of the mode by elementary
methods. For an alternative derivation of the mass see \cite{miemiec}.
These modes are responsible for the long-range gravitational
interactions on the brane and they reproduce the Schwarzschild
AdS(dS) black hole at linearized order on the brane.

\section*{Acknowledgments}
This paper is based on work with J. T. Liu and H. c. Ren and it
is supported in part by the US Department of Energy under grants
DE-FG02-91ER40651-TASKB.

\ifpreprintsty\else
\fi

\end{document}